\documentstyle[11pt,amssymb,epsf]{article}

\def\dalemb#1#2{{\vbox{\hrule height .#2pt
        \hbox{\vrule width.#2pt height#1pt \kern#1pt
                \vrule width.#2pt}
        \hrule height.#2pt}}}
\def\square{\mathord{\dalemb{6.8}{7}\hbox{\hskip1pt}}}

\def\0{{\sst{(0)}}}
\def\1{{\sst{(1)}}}
\def\2{{\sst{(2)}}}
\def\3{{\sst{(3)}}}
\def\4{{\sst{(4)}}}
\def\5{{\sst{(5)}}}
\def\6{{\sst{(6)}}}
\def\7{{\sst{(7)}}}
\def\8{{\sst{(8)}}}

\def\ep{\epsilon}
\def\td{\tilde}
\def\wtd{\widetilde}
\def\half{{\textstyle{1\over2}}}

\let\a=\alpha

\def\nn{\nonumber} \def\bd{\begin{document}} \def\ed{\end{document}}
\def\ds{\documentstyle} \let\fr=\frac \let\bl=\bigl \let\br=\bigr
\let\Br=\Bigr \let\Bl=\Bigl
\let\bm=\bibitem
\let\na=\nabla
\let\pa=\partial \let\ov=\overline
\newcommand{\be}{\begin{equation}}
\newcommand{\ee}{\end{equation}}
\def\ba{\begin{array}}
\def\ea{\end{array}}
\def\ft#1#2{{\textstyle{{\scriptstyle #1}\over {\scriptstyle #2}}}}
\def\fft#1#2{{#1 \over #2}}
\def\del{\partial}
\def\sst#1{{\scriptscriptstyle #1}}
\def\oneone{\rlap 1\mkern4mu{\rm l}}
\def\ie{{\it i.e.\ }}
\def\via{{\it via}}
\def\semi{{\ltimes}}
\def\str{{\rm str}}
\def\jm{{\rm j}}
\def\im{{\rm i}}
\def\bOmega{{{\bar\Omega}}}
\def\Qn{{{Q_{\sst{\rm N}}}}}
\def\tX{{{\wtd X}}}
\def\pc{{{\phantom{\chi}}}}

\def\mapright#1{\smash{\mathop{-\!\!\!-\!\!\!-\!\!\!-\!\!\!-\!\!\!
             \longrightarrow}\limits^{#1}}}
\def\maprightt#1#2{\smash{\mathop{-\!\!\!-\!\!\!-\!\!\!-\!\!\!-\!\!\!
             \longrightarrow}\limits^{#1}_{#2}}}

\newcommand{\ho}[1]{$\, ^{#1}$}
\newcommand{\hoch}[1]{$\, ^{#1}$}
\newcommand{\bea}{\begin{eqnarray}}
\newcommand{\eea}{\end{eqnarray}}
\newcommand{\ra}{\rightarrow}
\newcommand{\lra}{\longrightarrow}
\newcommand{\Lra}{\Leftrightarrow}
\newcommand{\ap}{\alpha^\prime}
\newcommand{\bp}{\tilde \beta^\prime}
\newcommand{\tr}{{\rm tr} }
\newcommand{\Tr}{{\rm Tr} }

\newcommand{\NP}{Nucl. Phys. }
\newcommand{\tamphys}{\it Center for Theoretical Physics\\
Texas A\&M University, College Station, Texas 77843}
\newcommand{\umich}{\it Michigan Center for Theoretical Physics\\
Randall Laboratory, Physics Department\\
University of Michigan, Ann Arbor, Michigan 48109}
\newcommand{\upenn}{\it Department of Physics and Astronomy\\
University of Pennsylvania, Philadelphia, Pennsylvania 19104}
\newcommand{\SISSA}{\it  SISSA-ISAS and INFN, Sezione di Trieste\\
Via Beirut 2-4, I-34013, Trieste, Italy}
\newcommand{\ihp}{\it Institut Henri Poincar\'e\\
  11 rue Pierre et Marie Curie, F 75231 Paris Cedex 05}

\newcommand{\imperial}{\it The Blackett Laboratory, Imperial College\hoch{2}\\
Prince Consort Road, London SW7 2BZ, UK}

\newcommand{\rutgers}{\it Department of Physics and Astronomy, Rutgers
University, Piscataway, NJ 08855}

\newcommand{\auth}{M. Cveti\v{c}\hoch{\dagger1}, M.J. Duff\hoch{\star2},
James T. Liu\hoch{\star2},  H. L\"u\hoch{\star2}, C.N. Pope\hoch{\ddagger3}
and K.S. Stelle\hoch{\#}}

\begin{document}
\begin{flushright}
\hfill{IHP-2000/03}\ \ \ {Imperial/TP/00-01/4}\ \ \ {RUNHETC-2000-47}
\ \ \ {SISSA-Ref.87/2000/EP}\\
\hfill{CTP TAMU-34/00}\ \ \ {MCTP-00-2}\ \ \
{UPR-911-T}\\
\hfill{November 2000}\ \ \ {hep-th/0011167}\\
\end{flushright}


\begin{center}
{ \large {\bf Randall-Sundrum Brane Tensions}}



\vspace{7pt}
\auth

\vspace{7pt}
{\hoch{\dagger}\hoch{\ddagger}\hoch{\#} \ihp}

\vspace{7pt}
{\hoch{\#}\imperial}

\vspace{7pt}
{\hoch{\dagger}\rutgers}

\vspace{7pt}
{\hoch{\dagger}\hoch{\ddagger}\hoch{\#}\SISSA}

\vspace{7pt}
{\hoch{\ddagger}\tamphys}

\vspace{7pt}
{\hoch{\star}\umich}

\vspace{7pt}
{\hoch{\dagger}\upenn}

\vspace{7pt}


\vspace{20pt}

\underline{ABSTRACT}
\end{center}

    We show that the singular sources in the energy-momentum tensor for
the Randall-Sundrum brane world, viewed as a solution of type IIB
supergravity, are composed of two elements.  One of these is a D3-brane
source with tension opposite in sign to the RS tension in five
dimensions; the other arises from patching two regions of flat
ten-dimensional spacetime.  This resolves an apparent discrepancy
between supersymmetry and the sign and magnitude of the RS tension.

{\vfill\leftline{}\vfill
\footnoterule
{\footnotesize \hoch{1} Research supported in part by DOE grant
DE-FG02-95ER40893 and NATO grant 976951. \vskip -12pt} \vskip 14pt
{\footnotesize \hoch{2} Research supported in part by DOE grant
DE-FG02-95ER40899. \vskip -12pt} \vskip 14pt
{\footnotesize  \hoch{3} Research supported in part by DOE
grant DE-FG03-95ER40917.\vskip  -12pt}}

\pagebreak
\setcounter{page}{1}

\section{Introduction}

    The original Randall-Sundrum brane-world model \cite{RS1} (which we will
subsequently refer to as Randall-Sundrum I), where our universe is viewed
as one of two 3-branes embedded at opposite ends of a patch of
five-dimensional anti-de Sitter space, was proposed as a mechanism
for naturally generating a large exponential hierarchy based on a small
extra dimension.  Subsequently, it was realized that the second brane
may be pushed off to the Cauchy horizon of the anti-de Sitter space,
yielding a model with a single 3-brane and a non-compact extra dimension
\cite{RS2}.  Much attention has been drawn to the fact that, contrary to
conventional Kaluza-Klein lore, this Randall-Sundrum II model is able to
trap gravity to the brane, even with an infinite fifth dimension.  This
localization of gravity to the brane occurs because the Randall-Sundrum
metric takes on a warped product form,
\begin{equation}
\label{ads5met}
ds_{5}^{2}=e^{-2K\, |z|}\, \eta_{\mu\nu}\, dx^\mu dx^\nu+dz^2\,,
\end{equation}
with a pronounced ``kink'' at the brane.\footnote{Such types of solutions
were first considered in $D=4$, $N=1$ supergravity in \cite{cgr}. 
For a review, see \cite{cs}}.

   Since the Randall-Sundrum geometry, given by the above metric,
is essentially that of two Poincar\'e patches of AdS$_5$ joined together
on a horospherical slice, recent attempts have been made to provide a more
stringy origin of the scenario and to relate it to Maldacena's AdS/CFT
conjecture \cite{Maldacena,Wittenads,GKP}.  Hints of this connection were
made in \cite{Verlinde,Gubser,Giddings,Hawking}, where the Randall-Sundrum
II model was presented as an anti-de Sitter bulk theory with a boundary not
at infinity but instead at some finite location.  By pushing the boundary
off to infinity, one finally decouples gravity and hence recovers the
standard AdS/CFT relation.  Further evidence for the equivalence of the
two scenarios was presented in \cite{Duffliu}, which demonstrated that
the $1/r^3$ correction the Newtonian potential in the Randall-Sundrum
background is reproduced exactly by the one-loop correction to the
graviton propagator in the corresponding CFT.

   Since AdS$_5$ arises as the near-horizon geometry of D3-branes in
the type IIB theory, it is natural to seek a D3-brane interpretation
for the Randall-Sundrum geometry.  This connection was made in
Refs.~\cite{clp99,clp00,deAlwis:2000qc,deAlwis:2000pr,dls}, which
makes use of a massive five-dimensional supergravity scenario
originating from an $S^5$ compactification of type IIB theory with a
breathing mode \cite{bremer}.  This result ensures that the
Randall-Sundrum scenario may be realized in a fully supersymmetric
context.  Despite some initial no-go theorems
\cite{Kallosh:2000tj,Behrndt} and difficulty in supersymmetrizing the
brane-world in a pure $N=2$ supergravity context \cite{ABN}, it has
subsequently been seen that these difficulties are overcome through
the incorporation of the breathing mode, and by appealing to the
higher-dimensional origin of the underlying gauged five-dimensional
supergravity \cite{dls,bkvp,sati}.

  Nevertheless, investigation of the supersymmetric Randall-Sundrum
realization \cite{dls} has yet to fully resolve the issue of brane
tension and the nature of the singular sources for the Randall-Sundrum
brane located at $z=0$.  It was pointed out in \cite{Kraus:1999it,dls}
that the ten-dimensional picture for a Randall-Sundrum II geometry
ought to be two segments of positive-energy D3-brane geometry, each
carrying $N$ units of charge, with $2N$ D3-branes sitting at the
patching location, causing the total charge to add up to zero.  It was
noted in \cite{dls} that if supersymmetry were to be preserved, the
$2N$ D3-branes at the patching location would have to have negative
tension.  However two puzzles arise with this picture of the
Randall-Sundrum geometry.  Firstly, as shown in \cite{Kraus:1999it},
the magnitude of the ``tension'' of the Randall-Sundrum brane \cite{RS1}
is greater than that of a corresponding stack of $2N$
D3-branes.\footnote{This was in fact initially used as an argument for
why the Randall-Sundrum brane could not be supersymmetric.}  Secondly,
even if the magnitudes of the ``tensions'' agreed, it is not obvious
how negative tension in $D=10$ could lead to positive tension in
$D=5$.

   The resolution to both of these puzzles comes from the realization
that the patching of the two stacks of positive tension D3-branes
involves more than simply the introduction of $2N$ negative-tension
D3-branes to absorb the charge.  In joining together two spacetime
patches on a curved boundary (which is what occurs in this
construction), one must be careful to introduce an appropriate source
of curvature in order to satisfy the Israel junction conditions.  By
careful examination of the Randall-Sundrum metric lifted to $D=10$, we
find that there are in fact two sources of such curvature.  The first
is of course the stack of the $2N$ negative tension D3-branes, and the
second arises from the $Z_2$ identification of transverse space (which
remains present even in the limit when the D3-branes are removed).
This dual-source nature of the Randall-Sundrum brane becomes more
apparent if the two sources are separated rather than superimposed.

   To study this, we begin in section 2 by considering the
Randall-Sundrum brane in $D=5$, and its lifting to $D=10$.  We find
that the singular energy-momentum sources in $D=10$ do not correspond
purely to a D3-brane, but instead they have a second contribution
coming from the $Z_2$ identification of flat space.  The picture
becomes clearer if the five-dimensional solution is generalised to
include the breathing mode, since then the ten-dimensional solution
describes actual D3-branes rather than merely their near-horizon
limit.  We therefore extend our discussion of the ten-dimensional
singular sources to cover this more general class of solution.  A
general feature both for these solutions, and for the pure
Randall-Sundrum limit, is that the positive-energy brane wall at $z=0$
turns out to be the sum of a negative-tension D3-brane superimposed on
a positive-energy singular source corresponding to making the $Z_2$
identification of flat spacetime.  This latter term outweighs the
negative contribution from the D3-brane, and so the net energy is
still positive in ten dimensions.  We then show how in detail how the
tension of the D3-brane agrees perfectly, both in magnitude and sign,
with the expectation based upon a careful application of the arguments
in \cite{Kraus:1999it}.

   In section 3, we identify and separate the D3-brane
$Z_2$-identification sources for
both the Randall-Sundrum I and II scenarios.  In section 4, we
explore the behavior of the massless graviton wave function for the case of
separated sources.  We end with conclusions and some speculations in
section 5, including a possible interpretation of the
$Z_2$-identification sources in terms of averaged arrays of
D7-branes.  This might give a relation to the F-theory picture
described in \cite{Chan:2000ms}.

\section{The brane tension in $D=10$}

    In this section, we shall consider the singular sources in $D=10$
for the Randall-Sundrum model itself (\ie two pieces of AdS$_5$ glued
together) and also a family of generalisations involving the massive
breathing-mode scalar field in $D=5$.  We shall postpone the
discussion of these generalisations until the next subsection, and
shall concentrate first, for simplicity, on the pure Randall-Sundrum
case, since this example already illustrates all the essential points.

\subsection{The singular sources for the Randall-Sundrum solution}

   In horospherical coordinates, the Randall-Sundrum solution takes
the form (\ref{ads5met}), which in the bulk is a solution of the
five-dimensional Einstein equation with a cosmological constant.  The
absolute-value symbol on the $z$ in the exponential implies that there
will be a delta-function in the curvature, and in fact the vielbein
components of the Ricci tensor are given by
\bea
R_{\mu\nu} &=& -4K^2\, \eta_{\mu\nu} + 2K\, \delta(z)\, \eta_{\mu\nu}
\,,\nn\\
R_{55} &=& -4K^2 + 8K\, \delta(z) \,.\label{adsric}
\eea
{}From this it follows that the Einstein tensor $G_{mn} \equiv R_{mn}-\ft12 R\,
\eta_{mn}$ is given by
\bea
G_{\mu\nu} &=&  6K^2\, \eta_{\mu\nu} -6K\,
\delta(z)\, \eta_{\mu\nu} \,,\nn\\
G_{55} &=&  6K^2\,.\label{adseinst}
\eea

   The structure of the singular terms in (\ref{adseinst}), and, in
particular, the absence of any singularity in $G_{55}$, is suggestive
of the energy-momentum tensor for a 3-brane source.  Note that the
Randall-Sundrum Lagrangian,
\begin{equation}
e^{-1}{\cal L}=R-\Lambda-\tau_{\rm RS}^\pc\, (g_{55})^{-1/2}\, \delta(z),
\end{equation}
yields the Einstein equation
\begin{equation}
\label{eq:rseins}
G_{MN}=-\ft12\Lambda\, \eta_{MN}-\ft12\tau_{\rm RS}^\pc\,
\eta_{\mu\nu}\, \delta_M^\mu\,
\delta_N^\nu\, (g_{55})^{-1/2}\, \delta(z).
\end{equation}
Comparing (\ref{eq:rseins}) with (\ref{adseinst}) then fixes the
Randall-Sundrum ``tension'' to be $\tau_{\rm RS}^\pc=12K$ and the
cosmological constant to be $\Lambda=-12K^2$ \cite{RS1,RS2}.

However, it should be borne in mind that there is no fundamental
3-brane as such in $D=5$, and so we should really reserve judgment
about how to interpret the singular terms and the meaning of
$\tau_{\rm RS}^\pc$ until we have lifted the
solution back to $D=10$ type IIB supergravity.  This is easily done;
the only non-vanishing ten-dimensional fields will be the metric and
the self-dual 5-form, given by
\bea
d\hat s_{10}^2 &=& ds_5^2 + m^{-2}\, d\Omega_5^2\,,\nn\\
\hat F_\5 &=& 4m \, \ep_\5 + 4\, m^{-4}\, \Omega_\5\,,\label{tenads}
\eea
where $K^2=m^2$, $\ep_5$ is the volume form of the metric $ds_\5^2$
given in (\ref{ads5met}), and $\Omega_\5$ is the volume form of the
unit $S^5$ metric $d\Omega_5^2$.  In fact we should choose
\be
m=  \left\{\matrix{ +K\,, & z>0\,, \cr
                    -K\,, & z<0\,, } \right. \label{mgrel}
\ee
in order to ensure continuity of the Killing spinors across the
junction at $z=0$ \cite{dls}.

   It is now a simple matter using (\ref{adsric}) to calculate the
Ricci tensor $\hat R_{AB}$, and hence the Einstein tensor $\hat
G_{AB}$, in $D=10$.  We find that the vielbein components are given by
\bea
\hat G_{\mu\nu} &=& -4K^2\, \eta_{\mu\nu}
-6K\, \delta(z)\, \eta_{\mu\nu} \,,\nn\\
\hat G_{55} &=& - 4K^2\,,\label{ads5ten}\\
\hat G_{ab} &=& 4K^2\, \delta_{ab} - 8K\, \delta(z)\, \delta_{ab}\,.\nn
\eea
Naively, by looking just at the $\hat G_{\mu\nu}$ and $\hat G_{55}$
components, we might be tempted to interpret the singularities as
being due to a D3-brane source with tension proportional to
\be
\td\sigma= 12K\,.\label{wrong}
\ee
However, it is evident from the fact that there are singular terms
also in the $S^5$
directions $\hat G_{ab}$ that we cannot simply attribute the
singularities to a D3-brane source, which would have the
structure
\bea
\hat T^{\rm{D3}}_{\mu\nu}
 &=& -\half\td\sigma\, \delta(z)\, \eta_{\mu\nu} \,,\nn\\
\hat T^{\rm{D3}}_{55} &=& 0\,,\label{d3source}\\
\hat T^{\rm{D3}}_{ab} &=& 0 \,.\nn
\eea

The precise relation between $\td\sigma$ and the D3-brane tension
$\tau_{\rm D3}^\pc$ may be determined by considering the $D=10$ supergravity
coupled D3-brane action
\begin{equation}
S_{10}=\int d^{10}x\sqrt{-\hat g}\, [\hat R+\cdots]
-\tau_{\rm D3}^\pc\int d^4\xi\sqrt{-\overline{g}}
+\mu_{D3}\int_{{\cal M}_4}\hat A_{[4]}
\end{equation}
(for vanishing gauge fields on the D3-brane) where $\overline{g}$ is the
induced metric on the brane.  In physical gauge, the resulting stress tensor
for a single D3-brane source has the form
\begin{equation}
\label{eq:d3src}
\hat T_{MN}^{\rm D3}=-\ft12\tau_{\rm D3}^\pc\, \eta_{\mu\nu}\, \delta_M^\mu
\, \delta_N^\nu\, \fft{\delta^6(\vec y-\vec y_0)}{\sqrt{\det\hat g_{ij}}},
\end{equation}
where $\hat g_{ij}$ is the metric in the space transverse to the brane,
which is located at $\vec y_0$.  To enforce the $S^5$ symmetry in
(\ref{tenads}), one must take a spherical distribution of branes.  For a
total of $N$ D3-branes in the above geometry, the stress tensor becomes
\begin{equation}
\hat T_{MN}^{\rm D3}=-\ft12N\, K^5\pi^{-3}\, \tau_{\rm D3}^\pc\,
\eta_{\mu\nu}\, \delta_M^\mu\,
\delta_N^\nu\, \delta(z),
\end{equation}
where $K^{-5}\, \pi^3$ is the volume of $S^5$.  From the
ten-dimensional metric in (\ref{tenads}), we therefore find that the
coefficient $\td\sigma$ in (\ref{d3source}) is related to the
fundamental D3-brane tension by $\td\sigma=N\, K^{5}\, \pi^{-3}\, \tau_{\rm
D3}^\pc$.

   In fact the singularities in the Einstein tensor (\ref{ads5ten})
can be understood as coming from two distinct sources.  One
contribution is indeed a fundamental D3-brane, while the
other is a contribution that would arise even in flat space, in the
absence of any D3-brane, as a result of having performed the $Z_2$
identification implied by the use of the absolute-value symbol on $z$
in (\ref{ads5met}).  It is easier to understand this second
contribution in the more general context of the breathing-mode
solutions that we shall discuss below.  However, in order to be able
to complete our present discussion for the AdS$_5$ solution we shall
just quote for now from some results that we shall derive later.  We
find that the ten-dimensional Einstein tensor that results from the
$Z_2$ identification of flat space has singularities that are supplied
by an energy-momentum tensor of the following structure:
\bea
\hat T^{Z_2}_{\mu\nu}
 &=& -\half\td\kappa\, \delta(z)\, \eta_{\mu\nu} \,,\nn\\
\hat T^{Z_2}_{55} &=& 0\,,\label{turtlesource}\\
\hat T^{Z_2}_{ab} &=& -\ft25 \td\kappa\, \delta(z)\, \delta_{ab} \,.\nn
\eea
It is the {\it sum} $\hat T^{\rm{D3}}_{AB} + \hat T^{Z_2}_{AB}$ that
should be matched to the singular terms in the Einstein tensor
(\ref{ads5ten}).  Clearly we shall have $\td\kappa=20K$, and so instead
of the D3-brane tension being given by (\ref{wrong}), it is actually
given by
\be
\td\sigma = - 8K\,.
\label{eq:d3t}
\ee
Not only is the correct tension smaller by a factor of $\fft23$ than the
``naive'' value \cite{Kraus:1999it}, it is also of the opposite sign!  Thus a
Randall-Sundrum domain wall in five dimensions that would
conventionally be described as having ``positive tension'' is actually
supported, from the ten-dimensional viewpoint, by a negative-tension
D3-brane source together with a superimposed source term associated
with the $Z_2$ identification of flat space.  The latter contribution
outweighs the former, so that the sign of the total $\hat T_{00}$
contribution is still positive in $D=10$.

At this point it is useful to revisit the tension argument of
\cite{Kraus:1999it}, which relates the $D=10$ D3-brane tension to
$D=5$ brane-world quantities.  As indicated in \cite{Kraus:1999it,dls}, the
supersymmetric Randall-Sundrum realization has the brane-world sitting between
two stacks of $N$ D3-branes.  The AdS curvature is then related to $N$
by \cite{Maldacena}
\begin{equation}
4\pi^3\, K^{-4}=N\, \tau_{\rm D3}^\pc.
\label{eq:adsc}
\end{equation}
Based on flux conservation, the Randall-Sundrum brane must then carry $-2N$
units of charge.  In order for this configuration to be supersymmetric,
however, this negative charge must be accompanied by negative tension.
Using the previously derived relation between $\tilde\sigma$ and
$\tau_{\rm D3}^\pc$, the corresponding tension of $2N$
negatively-charged D3-branes is then \cite{Kraus:1999it}
\begin{equation}
\tilde\sigma=-2N\, K^5\, \pi^{-3}\, \tau_{\rm D3}^\pc=-8K,
\end{equation}
which agrees with (\ref{eq:d3t}), and thus confirms the above split of $\hat
T_{MN}$ into a D3-brane source and the $Z_2$ identification of flat space.

   Although this result, especially the change of sign of the D3-brane
tension, might seem a little surprising, it is actually rather natural
from the ten-dimensional viewpoint.  It is rather easier to understand
in the context of the more general family of Randall-Sundrum type
solutions that make use of the breathing-mode scalar in five
dimensions, and it is to this subject that we now turn.

\subsection{Singular sources for the breathing-mode solutions}

   Here, we consider generalisations of the Randall-Sundrum solution,
namely certain 3-brane solutions in five dimensions that are supported
by a subset of the complete set of fields that come from the $S^5$
reduction of type IIB supergravity.  Specifically, the relevant fields
are the metric tensor and the breathing-mode scalar that parameterises
the overall volume of the 5-sphere.  The solutions were first obtained
in \cite{bremer}, and the idea of using them in the context of a
Randall-Sundrum II scenario was introduced in \cite{clp99}, and
studied further in \cite{clp00,deAlwis:2000qc}.  The reason for
considering the breathing-mode scalar was that it is a massive field
(lying outside the massless supergravity multiplet that is commonly
considered), and its potential has a minimum rather than a maximum.
It was shown in \cite{clp00} that although it still cannot give rise
to a smooth gravity-trapping domain wall, it can give an acceptable
singular wall of the Randall-Sundrum II type if two segments of the
solution are appropriately patched with a delta-function in the
curvature at the ``join.''  The same bulk solution was then used in
\cite{dls} for the construction of a Randall-Sundrum I type of
scenario, with two singular branes, one of positive and the other of
negative tension.

    To study the brane tensions in detail, we first need to review the
pertinent aspects of the $S^5$ reduction of type IIB supergravity to
obtain the theory in $D=5$ with the breathing mode, and to find the
brane solution in five dimensions.  These results are taken from
\cite{bremer}.

       The only non-vanishing fields in the type IIB theory are the
metric $\hat g_{MN}$ and the self-dual 5-form $\hat F_\5$.  The type
IIB field equations $\hat R_{MN} = \ft1{96} \hat F^2_{MN}$, ${\hat
*\hat F_\5}=\hat F_\5$ and $d\hat F_\5=0$ are satisfied by the Ansatz
\bea
d\hat s_{10}^2 &=& e^{\fft12\sqrt{\fft53}\varphi}\, ds_5^2 +
e^{-\fft12\sqrt{\fft35}\varphi}\, ds^2(S^5)\,,\nn\\
\hat F_\5 &=& 4m \, e^{2\sqrt{\fft53}\varphi}\, \ep_\5 + 4m\,
\ep(S^5)\,,\label{10ans}
\eea
provided that the five-dimensional fields $ds_5^2$ and $\varphi$
satisfy the equations of motion following from the Lagrangian
\be
e^{-1}\, {\cal L}_5 = R -\ft12 (\del\varphi)^2 - 8m^2\,
e^{2\sqrt{\fft53}\varphi}\, + e^{\fft4{\sqrt{15}}\varphi}\, R_5\,.
\ee
Here $\ep_\5$ is the volume form of $ds_5^2$, $\ep(S^5)$ is the volume
form of the metric $ds^2(S^5)$ on the 5-sphere, which has (constant)
Ricci scalar $R_5$, and $m$ is another constant.  It is useful to note
that the vielbein components of the ten-dimensional Ricci tensor for
the metric $d\hat s_{10}^2$ in (\ref{10ans}) are given by
\bea
\hat R_{mn} &=& e^{-\fft12\sqrt{\fft53}\varphi}\, \Big( R_{mn} - \ft14
\sqrt{\ft53} \square\varphi\, \eta_{mn} - \ft12 \nabla_m\varphi\,
\nabla_n\varphi\Big)\,,\nn\\
\hat R_{ab} &=& e^{\fft12\sqrt{\fft35}\varphi}\,  R_{ab} + \ft14
\sqrt{\ft35}\,  e^{-\fft12 \sqrt{\fft53}\varphi}\,
\square\varphi\, \delta_{ab}\,,\label{ricred}\\
\hat R_{ma} &=& 0\,,\nn
\eea
where the index $m$ lies in the five-dimensional spacetime, and $a$
lies in $S^5$.

     The relevant domain-wall solution of the five-dimensional equations of
motion is given by \cite{bremer}
\bea
ds_5^2 &=& e^{2A}\, dx^\mu \, dx_\mu + e^{-8A}\, dz^2\,,\nn\\
e^{-\fft{7}{\sqrt{15}}\, \varphi} &=& H= e^{-\fft{7}{\sqrt{15}}\, \varphi_0}
+ k\, |z|\,,\label{solution}\\
e^{4A} &=& \td b_1\, H^{\fft27} + \td b_2\, H^{\fft57}\,,\nn
\eea
where $\td b_1 = \mp 28m/(3k)$ and $\td b_2= \pm 14 \sqrt{5 R_5} \, /(15k)$.
Of course, by placing the absolute-value symbol around the
coordinate $z$ we have caused delta-functions to appear in the
expressions for the Ricci tensor and $\square\varphi$, and so while
(\ref{solution}) solves the five-dimensional equations of motion
exactly when $z\ne0$, there will be a need for singular source terms
on the domain wall itself.

   Solutions exist for all four independent choices of signs for $\td b_1$ and
$\td b_2$, but the ones that are relevant here have $\td b_1\, \td b_2
<0$.  The AdS$_5$ limit is reached by sending $k$ to zero, with the
constant $\varphi_0$ given by
\be
e^{\fft6{\sqrt{15}}\varphi_0}= \fft{R_5}{20 m^2}\,.
\ee
The solution (\ref{solution}) in this limit has
\be
e^{4A}=4m\, e^{\fft5{\sqrt{15}}\varphi_0}\,
(z_0-|z|)\,,\qquad \varphi=\varphi_0\,,
\ee
and can be seen, after a simple coordinate transformation, to be
equivalent to the AdS$_5$ solution (\ref{ads5met}).
The constant $K$ appearing in (\ref{ads5met}) and (\ref{adsric}) is given by
\be
K= m\, \Big(\fft{R_5}{20 m^2}\Big)^{5/6}\,.
\ee

   Since our purpose is to study the precise nature of the singular
sources, we shall not need to present the ``regular'' terms in what
follows, since it is already established from the results of
\cite{bremer} that the five-dimensional and ten-dimensional equations
are satisfied in the bulk.  We shall therefore generally simply
represent the regular terms by ``Reg'' in the equations.  A
straightforward calculation from (\ref{solution}) shows that
$\square\varphi$ and the vielbein components of the Ricci tensor
take the forms \cite{dls}
\bea
\square\varphi &=& -\fft{2\sqrt{15}}{7}\, c^{-7}\, k \, (\td b_1\, c^2 +
\td b_2\, c^5)\,(g_{55})^{-1/2}\, \delta(z) + \hbox{Reg}\,,\nn\\
R_{\mu\nu} &=& -\ft1{14} k\, (2\td b_1\, c^{-5} + 5\td b_2\, c^{-2})
\,(g_{55})^{-1/2} \, \delta(z)\, \eta_{\mu\nu} +
\hbox{Reg}\,,\label{phiricdel}\\
R_{55} &=& -\ft27 k\,  (2\td b_1\, c^{-5} + 5\td b_2\, c^{-2})
\, (g_{55})^{-1/2} \, \delta(z) + \hbox{Reg}\,,\nn
\eea
where for convenience we have defined the constant $c$ by
\be
c\equiv e^{-\fft1{\sqrt{15}}\, \varphi_0}\,.\label{cdef}
\ee
Note that the five-dimensional Einstein tensor is therefore given by
\bea
G_{\mu\nu}&=& \ft3{14} k\, (2\td b_1\, c^{-5} + 5\td b_2\, c^{-2})
\, (g_{55})^{-1/2} \, \delta(z)\, \eta_{\mu\nu} + \hbox{Reg}\,,\nn\\
G_{55} &=& 0 + \hbox{Reg}\,.\label{d5einst}
\eea

   These singular terms have the general structure one would expect of
a 3-brane source in $D=5$.  However, as in our previous discussion for
the AdS$_5$ case, the absence of a fundamental 3-brane in $D=5$ means
that we should postpone interpreting the singular terms until we have
lifted the solution back to $D=10$.  It is known that the bulk
solution (\ref{solution}) oxidises to give a standard D3-brane
supergravity solution in $D=10$ \cite{bremer}, namely
\be
d\hat s_{10}^2 = \wtd H(r)^{-\fft12}\, d\td x^\mu\, d\td
x_\mu + \wtd H(r)^{\fft12}\, (dr^2 + r^2\, d\Omega_5^2)\,,\label{d3sol}
\ee
where
\be
\wtd H(r) \equiv 1 + \fft{\td k}{r^4}\,,\label{thdef}
\ee
and
\be
\td x^\mu = \td b_2^{1/2}\, x^\mu\,,\qquad r^4 = (20/R_5)^2\,
H^{\fft37} - \td k\,,\qquad \td k = -(20/R_5)^2\, \fft{\td b_1}{\td
b_2}\,,\label{redefs}
\ee
as can be seen by substituting (\ref{solution}) into (\ref{10ans}).
One might be tempted to expect that after lifting the source terms to
$D=10$, they would become precisely those for the D3-brane.  As we
shall now show, this is not the case.

    It is a straightforward matter to lift the five-dimensional
expressions (\ref{phiricdel}) back to $D=10$, using the reduction
Ansatz (\ref{10ans}).  In particular, using (\ref{ricred}) we can
calculate the vielbein components of the ten-dimensional Einstein
tensor, for the oxidation of the domain-wall solution (\ref{solution})
to $D=10$, finding
\bea
\hat G_{\mu\nu} &=&  \ft3{14}(2\td b_1 + 5\td b_2\, c^3)\, c^{-15/4}\, k
\, (\hat g_{55})^{-1/2} \, \delta(z)\, \eta_{\mu\nu} + \hbox{Reg}\,,\nn\\
\hat G_{55} &=& 0 + \hbox{Reg}\,,\label{hatdel}\\
\hat G_{ab} &=& \ft67\td b_2\,c^{-3/4}\, k\,
(\hat g_{55})^{-1/2} \, \delta(z)\, \delta_{ab} + \hbox{Reg}\,.\nn
\eea
Clearly this is not of the form of a pure 3-brane source in $D=10$,
which would have the structure given in (\ref{d3source}).

   As we have already foreshadowed, the resolution of the puzzle is that
the singular terms we are seeing in (\ref{hatdel}) are actually
accounted for by the sum of two quite distinct singular sources.  One
of these is a D3-brane source, and the other describes the curvature
singularities that result from identifying two patches of flat
spacetime under $Z_2$.  In fact the situation is made unnecessarily
confusing by having these two sources located at the same value of
the coordinate $z$ (or $r$) transverse to the brane, and later on we shall
find it convenient to separate the two contributions and locate them
at different values of $r$.

    To recognise how the singular sources for (\ref{hatdel}) separate
into the two contributions we can either first isolate the D3-brane
contribution, after which the remainder must be the term describing the
$Z_2$ identification of flat space, or else we can first isolate the
contribution from the $Z_2$ identification.  Although it is perhaps
logically more natural to follow the former approach, calculationally
it seems to be simpler first to calculate the curvature singularities
for $Z_2$-identified flat space, and then after extracting this part
from (\ref{hatdel}); the remainder will be the D3-brane contribution.

   To calculate the Ricci curvature for $Z_2$-identified flat space,
it is simplest to perform the computation in $D=5$, and then lift the
results back to $D=10$.  We begin by picking a radius $r=r_0$ in
(\ref{d3sol}), and then taking the flat-space metric that continuously
matches at this radius, namely
\be
d\hat s_{10}^2 = \wtd H(r_0)^{-\fft12}\, d\td x^\mu\, d\td
x_\mu + \wtd H(r_0)^{\fft12}\, (dr^2 + r^2\, d\Omega_5^2)\,,\label{flatsol0}
\ee
We now reduce this to $D=5$, using the reduction Ansatz
(\ref{10ans}).  At the same time, we make the following redefinitions
\be
r_0^4 + \td k = c^3\,,\qquad r^4 = r_0^4 + k\, |\xi|\,.
\ee
Thus by using the $\xi$ coordinate, we have arranged that as $\xi$ passes
from negative to positive, $r$ approaches $r_0$, and then retraces its
steps.  This introduces a ``join'' at $r=r_0$.
In terms of $z$, the $D=5$ metric becomes
\bea
ds_5^2 &=& r_0^{-\fft43}\, c\, (r_0^4+ k\, |\xi|)^{\fft56}\,
dx^\mu\, dx_\mu + \ft1{16}\, k^2\, r_0^{-\fft{16}{3}}\, c^4\,
 (r_0^4+ k\, |\xi|)^{-\fft23}\, d\xi^2\,,\nn\\
e^{-\sqrt{\fft35}\varphi} &=& r_0^{-4}\, c^3\, (r_0^4+ k\, |\xi|)\,.
\eea
We then find that
\bea
R_{\mu\nu} &=& \ft16 \td\kappa\,c^{-\fft54}\,
        (g_{55})^{-1/2}\, \delta(\xi)\,\eta_{\mu\nu} + \hbox{Reg}\,,\nn\\
R_{55} &=&  \ft{2}{3}\td \kappa \,c^{-\fft54}\,
(g_{55})^{-1/2}\, \delta(\xi) + \hbox{Reg}\,,\label{d5turtle}\\
\ft14\sqrt{\ft53}\,
\square\varphi &=&  \ft16 \td \kappa\,c^{-\fft54}\,
(g_{55})^{-1/2}\, \delta(\xi) + \hbox{Reg}\,,\nn
\eea
where
\be
\td\kappa \equiv -20\, c^{-\fft{3}{4}}\,.
\ee
Lifting to $D=10$, we find that the Einstein tensor for
$Z_2$-identified flat space has the following vielbein components:
\bea
\hat G_{\mu\nu} &=& -\half\td\kappa\,
(\hat g_{55})^{-1/2}\, \delta(\xi)\, \eta_{\mu\nu} \,,\nn\\
\hat G_{55} &=& 0 \,,\label{d10turtle} \\
\hat G_{ab}  &=& -\ft25\td\kappa\,
(\hat g_{55})^{-1/2}\, \delta(\xi)\, \delta_{ab}
\,.\nn
\eea
Since this is not a brane-source term it is inappropriate to speak of
a ``tension'' associated with this contribution.  We shall merely
describe it as having positive energy if $\td\kappa$ is positive, and
{\it vice versa}.

   Using this result, we can now make the unique decomposition of the
singular terms in the ten-dimensional Einstein tensor (\ref{hatdel})
into the sum of a D3-brane source (\ref{d3source}) and the identified
flat-space source (\ref{d10turtle}).  Thus we find
\be
\td\sigma = -  \fft{6k}{7}\, c^{-15/4}\, \td b_1\,,\qquad
\td\kappa =  -\fft{15 k}{7}\, c^{-3/4}\, \td b_2\,.\label{sigkap}
\ee
Using $\td b_1 = 28m/(3k)$ allows us to rewrite the D3-brane tension as
$\td\sigma=-8mc^{-15/4}$.  Now, in deriving the fundamental D3-brane
tension from (\ref{eq:d3src}), one has to account for the complete
Ansatz (\ref{10ans}) when determining the volume of $S^5$.  The
resulting relation is
\begin{equation}
\td\sigma(\hbox{single D3-brane}) = \left(R_5\over20\right)^{5/2}
\pi^{-3}\, c^{-15/4}\, \tau_{\rm D3}^\pc.
\end{equation}
At the same time, the AdS relation, (\ref{eq:adsc}), is modified to%
\footnote{This relation may be derived by comparing the factor $\td k$ in
the D3-brane harmonic function, (\ref{thdef}), with that demanded by
charge quantization.}
\begin{equation}
4\pi^3\, m\, \left(R_5\over20\right)^{-5/2}=N\, \tau_{\rm D3}^\pc.
\end{equation}
Thus one finds simply
\begin{equation}
\td\sigma=-2N\, \td\sigma(\hbox{single D3-brane}),
\end{equation}
which once again corresponds to having a source of $2N$ negative
tension D3-branes.

It is clear that the ``tension'' $\td\kappa$ arising from the $Z_2$
identification cannot be identified with any conventional brane source.
After all, a $p$-brane with tension $\tau_p$ would yield a contribution
to the $D=10$ stress tensor of the form
\begin{equation}
\hat T_{MN}^p=-\half\tau_p\, \eta_{\mu\nu}\, \delta_M^\mu\, \delta_N^\nu
\, \fft{\delta^{9-p}(\vec y-\vec y_0)}{\sqrt{\det\hat g_{ij}}},
\end{equation}
where $\hat g_{ij}$ is the metric transverse to the brane.  Unlike
(\ref{d10turtle}), this $p$-brane contribution is isotropic in all
directions longitudinal to the brane.

\section{Separating the singular sources}

     Having seen that the singular terms in the Einstein tensor in
$D=10$ are actually matched by two quite distinct sources, namely a
D3-brane and the term associated with the $Z_2$ identification of flat
space, we see that it now becomes natural to dissociate the two
terms from one another, by placing them at different values of $r$ in
the ten-dimensional solution.  We shall discuss this first for the
case of a Randall-Sundrum II type scenario, where there is just a
single domain wall in five dimensions, at $z=0$.  After doing this, we
shall then discuss the case of a Randall-Sundrum I model, where a
seond domain wall is introduced at $z=L$, by identifying the
coordinate values $z=+L$ and $z=-L$.

\subsection{Ten-dimensional sources for a single domain wall}

    It can be seen from (\ref{redefs}) that as $z$ ranges from $z=0$
on the five-dimensional domain wall to $z=\pm\infty$ on the Cauchy
horizon, the radial coordinate $r$ in the ten-dimensional D3-brane
solution ranges from some fixed value $r>0$ outside the horizon to
$r=0$, on the D3-brane horizon.  We shall now deform the geometry so
that the $S^5$ spherical shell of D3-brane source and the
$Z_2$-identification source are located at two different values of
$r$.  Specifically, as $r$ increases from the horizon at $r=0$, the
D3-brane source will be encountered first, at some radius $r=r_1$.  We
saw by direct calculation that the D3-brane tension was coming out to
be negative, and in fact this is very natural.  It has the effect of
``turning off'' the D3-brane charge for values of $r$ that lie outside
$r=r_1$, so that for $r>r_1$ the D3-brane solution (\ref{d3sol}) is
replaced by a flat spacetime.  Of course the metrics must match
continuously at $r=r_1$, and so we shall have that for $r>r_1$, the
flat metric is given by
\be
d\hat s_{10}^2 = \wtd H(r_1)^{-\fft12}\, d\td x^\mu\, d\td
x_\mu + \wtd H(r_1)^{\fft12}\, (dr^2 + r^2\, d\Omega_5^2)\,,\label{flatsol}
\ee
It should be remarked that one can explicitly check, using the
relations given in (\ref{redefs}), that the D3-brane tension $\sigma$
that we calculated in (\ref{sigkap}) is precisely the correct value to
``turn off'' the D3-brane (\ref{d3sol}) for $r$ lying outside the
radius $r_1$ at which we have now chosen to locate the D3-brane source.

    Proceeding outwards, we now choose a radius $r_0>r_1$ at which the
$Z_2$ identification of the flat spacetime (\ref{flatsol}) will be
performed.  This implies a singular source term of the form
(\ref{d10turtle}), with positive energy.  The ten-dimensional solution
is depicted in Figure 1.

\begin{figure}[ht]
\leavevmode\centering
\epsfbox{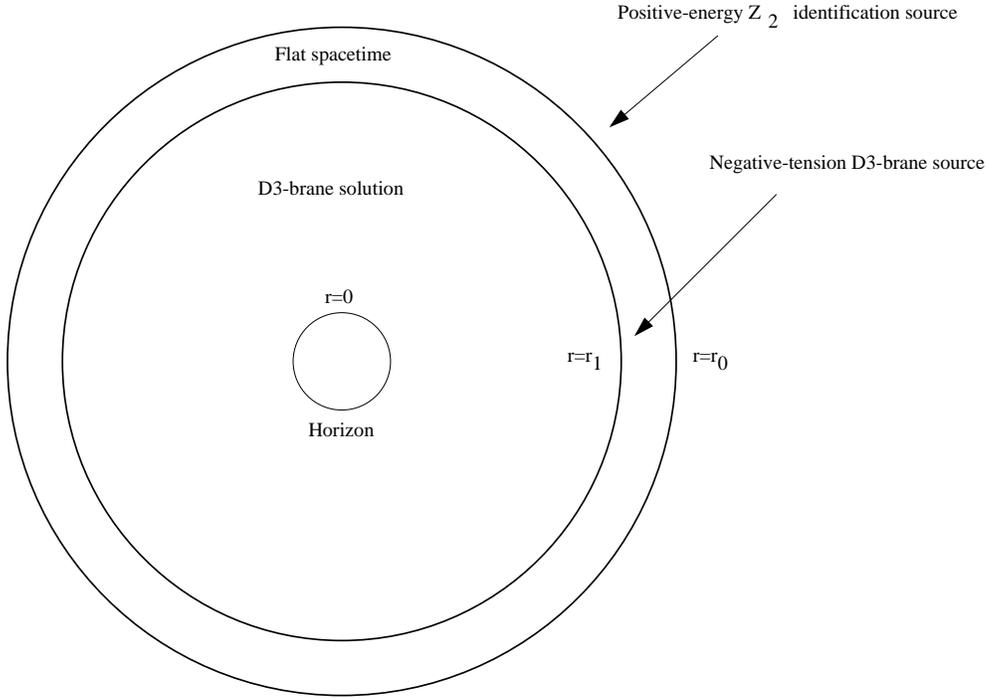}
\caption{The ten-dimensional configuration for the RS II brane}
\end{figure}

\subsection{Ten-dimensional sources for two domain walls}

    The situation described above is for a single domain wall in five
dimensions, corresponding to a model of the Randall-Sundrum II type.
We may also consider the situation where a second domain wall is
introduced at $z=L$, by identifying $z=L$ with $z=-L$.  This is
conventionally described, in five dimensions, as having ``negative
tension.''  Again, in ten dimensions we shall separate the singular
terms associated with this second wall into a D3-brane source and a
$Z_2$-identification source, located at two different values of $r$.
Since its five-dimensional ``tension'' is of the opposite sign to that
of the wall at $z=0$, we know from our earlier results that the
associated D3-brane source in $D=10$ will have positive tension.  The
second wall lies at a smaller radius $r$ than the original one.  Thus
we shall now have a total of four radii at which singular sources are
located, which we take to be in the order $r_3<r_2<r_1<r_0$.  The
disposition of sources will be as follows.  At $r=r_0$ is the
$Z_2$-identification source for the $z=0$ domain wall, and at $r=r_1$
is the negative-tension D3-brane source for the $z=0$ domain wall,
exactly as in the previous subsection.  Moving inwards, we next
encounter a positive-tension D3-brane source for the $z=L$ domain
wall, at $r=r_2$.  Finally, we encounter a $Z_2$-identification source
at $r=r_3$, which has negative energy.

    The region $r_2\le r \le r_1$ has the form of a solid annulus of
the D3-brane solution (\ref{d3sol}).  Outside this, for $r_1\le r\le
r_0$, and inside it, for $r_3\le r\le r_2$, are solid annuli of flat
spacetime.  The ten-dimensional spacetime is depicted in Figure 2.

\begin{figure}[ht]
\leavevmode\centering
\epsfbox{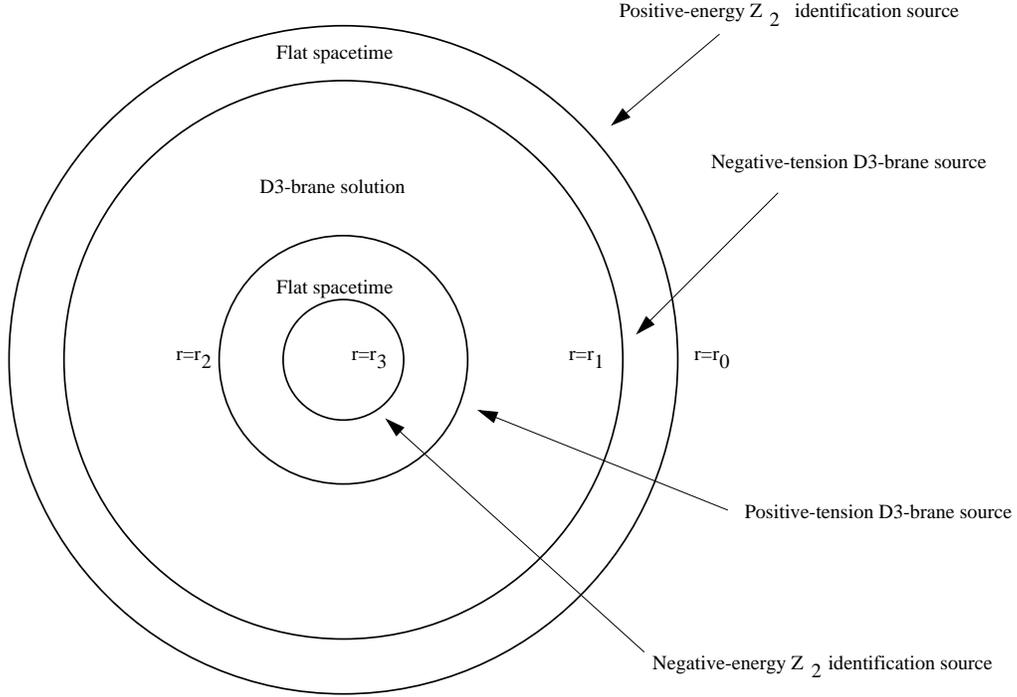}
\caption{The ten-dimensional configuration for the RS I branes}
\end{figure}

    In this situation with two domain walls, the effect of the $Z_2$
identification is as follows.  The coordinate $z$ can be thought of as
a coordinate on the circle $S^1$, with $-L\le z\le L$, which is then
factored by $Z_2$ in the style of Horava and Witten.  At the same
time, as one crosses from negative to positive $z$ the sign of the
parameter $m$ in the expression for $\hat F_\5$ in (\ref{10ans})
reverses, corresponding to a reversal of the orientation of the
compactifying 5-sphere.  Thus the ten-dimensional theory is reduced on
the orbifolded internal manifold $(S^1\times S^5)/Z_2$.

\section{The wave function for the massless graviton mode}

    In the conformally-flat frame
\be
ds_5^2 = e^{2A(z)}\, (dx^\mu\, dx_\mu + dz^2)\,,\label{conf}
\ee
the massless wave function is just given by
\be
\psi(z) = e^{\fft32 A(z)}\,.\label{wf}
\ee

    We can study this explicitly for the case where we take the pure
AdS$_5$ limit.  In fact it is easier then to derive the result
directly, rather than to get it by explicitly taking the singular
$k\longrightarrow 0$ limit.  But we still want to separate the
D3-brane and $Z_2$-identification sources.  Accordingly, we consider
the ten-dimensional metrics in two regimes:
\bea
\underline{ r\le r_1}: && d\hat s_{10}^2 = \fft{r^2}{r_1^2} \,
dx^\mu\, dx_\mu + \fft{r_1^2}{r^2}\, dr^2 + r_1^2\, d\Omega_5^2
\,,\nn\\
\underline{ r\ge r_1}: && d\hat s_{10}^2 =
dx^\mu\, dx_\mu +  dr^2 + r^2\, d\Omega_5^2
\,.
\eea
This is the AdS$_5$ solution, for $r\le r_1$, obtained by taking the
near-horizon limit in which the ``1'' is dropped in the harmonic
function $\wtd H(r) = 1 + \td k/r^4$, joined on continuously
to pure flat space for $r\ge r_1$.

    We now reduce to $D=5$, using the standard reduction Ansatz
\be
d\hat s_{10}^2 = e^{2\a\varphi}\, ds_5^2 + e^{-\fft65
\a\varphi}\, d\Omega_5^2\,.
\ee
Thus for the two regimes we have
\bea
\underline{ r\le r_1}: && ds_5^2 =  \fft{r^2}{r_1^2} \,
dx^\mu\, dx_\mu + \fft{r_1^2}{r^2}\, dr^2\,,\qquad \varphi=0\,,\nn\\
\underline{ r\ge r_1}: && ds_5^2 = \Big(\fft{r}{r_1}\Big)^{\fft{10}{3}}
\, (dx^\mu \, dx_\mu + dr^2)\,,\qquad e^{-\fft65\a\, \varphi} =
\fft{r^2}{r_1^2}\,.
\eea

   We now change to the $z$ coordinate, which transforms the metric
into the conformally-flat form (\ref{conf}).  In the two regimes we
therefore have
\bea
\underline{ r\le r_1}: &&  \fft{r_1}{r} = 1 + \fft{z-z_1}{r_1}\,,
\nn\\
\underline{ r\ge r_1}: &&  \fft{r}{r_1} = 1 + \fft{z_1-z}{r_1}\,,
\eea
where we have chosen the constants of integration for convenience, and
so that the join ($r=r_1$) occurs at $z=z_1$.

    We now choose a $Z_2$ identification at $r=r_0>r_1$.  We shall
take this to be at $z=0$, so now we have
\bea
\underline{ r\le r_1}: &&  \fft{r_1}{r} = 1 + \fft{|z|-z_1}{r_1}\,,
\nn\\
\underline{ r_1\le r\le r_0}: &&  \fft{r}{r_1} = 1 + \fft{z_1-|z|}{r_1}\,,
\eea
The upper equation corresponds to $|z|\ge z_1$, while the lower
corresponds to $|z|\le z_1$.  (We assume, without loss of generality,
that $z_1$ is positive.)  Note that we shall have
\be
z_1 = r_0-r_1\,.
\ee
To summarise, we now have a positive-energy source at $z=0$,
corresponding to $r=r_0$, which results from the $Z_2$ identification,
and a pair of negative-tension D3-brane sources at $z=\pm z_1$,
corresponding to $r=r_1<r_0$.  As the coordinate $z$ goes to the
Cauchy horizons at $\pm\infty$, the radial coordinate $r$ tends to zero.

    We see from
(\ref{wf}) that the wavefunction in the two regimes is given by
\bea
\underline{ |z| \ge z_1}: && \psi(z) = a\, e^{\fft32 A(z)} =
a\, \Big[1 + \fft{|z|-z_1}{r_1} \Big]^{-\fft32}\,,\nn\\
\underline{ |z|\le z_1}: && \psi(z) = a\, e^{\fft32 A(z)} =
a\, \Big[1 + \fft{z_1-|z|}{r_1}\Big]^{\fft52}\,,
\eea
where $a$ is a normalisation constant.  It can be determined by
requiring that $\int_{-\infty}^\infty dz\, |\psi|^2 =1$, and this
implies
\be
a = \fft{\sqrt 3\, r_1^{5/2}}{\sqrt{r_0^6 + 2 r_1^6}}\,.
\ee

   Expressed back in terms of the $r$ coordinate, the normalised
wavefunction becomes
\bea
\underline{0\le r \le r_1}: && \psi =
\fft{\sqrt3\, r_1\, r^{3/2}}{\sqrt{r_0^6 + 2 r_1^6}}\,,\nn\\
\underline{r_1\le r\le r_0}: && \psi =
\fft{\sqrt3\,  r^{5/2}}{\sqrt{r_0^6 + 2 r_1^6}}\,.
\eea
{}From this it follows that the wavefunction concentrates around the
$Z_2$-reflection point at $r=r_0$, even more strongly than in the
Randall-Sundrum case where $r_1=r_0$.  Conversely, it tends to be
suppressed on the D3-branes at $r=r_1$.

\section{Conclusion and speculations}

   In this paper, we have investigated the lifting of the pure
Randall-Sundrum AdS$_5$ solution, and its breathing-mode
generalisations, from $D=5$ to $D=10$.  Since the five-dimensional
breathing-mode solutions are nothing but D3-brane solutions when
lifted to $D=10$, this procedure provides a natural mechanism for
interpreting the singular sources in terms of ten-dimensional
D3-brane sources.  Likewise, the pure Randall-Sundrum AdS$_5$ limit
corresponds to taking the near-horizon limit of the D3-brane in ten
dimensions.

    After lifting the solutions, we find that the singular sources are
not accounted for purely by D3-brane sources in ten dimensions.  In
fact the singular terms in the energy-momentum tensor can be split into
the sum of a standard type of D3-brane source, plus a second term that
would be present even in the absence of any D3-branes, which results
from performing the $Z_2$ identification of flat spacetime.  Having
disentangled the two contributions, we were then able to show that the
D3-brane part has exactly the correct magnitude, and sign, to account
for the expected quantity $(-2N)$ of D3-brane charge that should be
present at $z=0$ in order to counterbalance the presence of $N$
D3-branes placed to the left and to the right of the brane wall.  In
other words, the puzzles that had been raised in \cite{Kraus:1999it},
which seemed to indicate a mismatch of brane charges, are now
fully resolved.

    An element of the resolution, the necessity for which was not
highlighted in \cite{Kraus:1999it}, is that not only is the magnitude
of the D3-brane charge different from the ``naive'' value that one
might be tempted to read off in $D=5$, but also the sign is reversed.
In particular, this means that what is customarily called a ``positive
tension brane'' in $D=5$ actually corresponds to a negative-tension
D3-brane in $D=10$.  This does not, however, mean that the sign of the
energy is reversed on lifting the solution from $D=5$ to $D=10$, since
the negative D3-brane contribution to the energy in $D=10$ is actually
outweighed by a larger positive contribution associated with the $Z_2$
identification of flat space.

   Having seen that the singular sources in $D=10$ have two
different kinds of contribution, it becomes natural to separate the
two, and allow them to be located at two different values of the
radial coordinate $r$ in the space transverse to the D3-brane.  In
fact it would seem now to be {\it unnatural} to insist that the two
contributions be superposed at the same spatial location, since the
position of the D3-brane is a modulus of the solution.  Furthermore,
the solutions will be supersymmetric for any choice of this modulus.

    Now that the five-dimensional ``positive-tension Randall-Sundrum
brane'' is recognised as corresponding to a negative-tension D3-brane
in $D=10$, the whole question of the possible instability of the model
raises its head again.  Also, one may wonder what happens concerning
the trapping of gravity.  These are two rather different questions,
and the latter seems to be an easier one to address.  We made a simple
analysis of the gravity trapping, by studying the form of the
wave-function for the massless four-dimensional graviton in the
Randall-Sundrum limit, after having first separated the D3-brane and
$Z_2$-identification sources as described above.  We found that the
wavefunction tends to peak on the positive-energy $Z_2$-identification
surface, and to be suppressed on the negative-tension D3-brane.
However, if the two singular surfaces are not too far separated, then
the binding effect on the $Z_2$-identification surface extends far
enough to encompass the D3-brane region too.  In this sense, gravity
is still trapped in the vicinity of the two surfaces.

   The question of stability, alluded to above, is less clear cut.  On
the one hand, the full one-parameter family of solutions with
arbitrary placement of the D3-brane source is supersymmetric.  This
would seem to indicate that no instabilities can be present.
Moreover, the full field theory clearly has a Bogomolnyi'i bound for
the energy.  On the other hand, a negative-tension D3-brane considered
in isolation would appear to have ``ballooning modes'' exploiting the
fact that the energy is decreased if the area increases.  The
resolution of the apparent conflict between these two viewpoints
remains unclear.

   So far, we have refrained from speculating about any physical
interpretation for the $Z_2$-identification singular sources.  We have
treated them for now as arising via some {\it deus ex machina},
without attempting to give the associated surfaces any dynamical
interpretation.\footnote{One might be tempted to view the
$Z_2$-identification energy-momentum sources as being provided by a
mechanism in the spirit of the turtles featured in the opening chapter
of Stephen Hawking's {\it A Brief History of Time.}  The universe,
according to an elderly lady attending a talk by a distinguished
cosmologist, is supported on the back of a giant turtle.  Before the
cosmologist could ask the obvious question, the lady quickly added
``and it's turtles all the way down.''}

    A more serious attempt at providing an interpretation for the
$Z_2$-identification sources might be the following.  From a type IIB
perspective, what is needed is an object that carries no five-form
charge, but that nevertheless folds up the transverse space.  While we
are unable to identify such a source in the maximally symmetric
AdS$_5\times S^5$ case, we can speculate on a possible interpretation
in a case where where there is a reduced symmetry in the transverse
space.  In such a circumstance, it seems that D7-branes may serve as a
source for singular $Z_2$-identification terms in the
energy-momentum tensor.  An examination of (\ref{d10turtle}) suggests
that if the assumption of spherical symmetry were relaxed, one could
wrap a set of D7-branes around a 4-cycle of the 5-dimensional compact
space.  The D7-branes would then contribute with a strength
$-\fft12\td\kappa$ to four out of the five components of $\hat
T_{ab}$, so that the factor of $-\fft25\td\kappa$ in (\ref{d10turtle})
results from `averaging' over all five internal directions.  For a
single wrapped D7-brane, the effective tension is given by%
\footnote{When the internal space is no longer $S^5$, the Ricci scalar $R_5$
would more properly be replaced by the average curvature of the space.}
\begin{equation}
\td\kappa(\hbox{single D7-brane}) = \left(\fft{R_5}{20}\right)^{5/2}
\, \pi^{-3}\, c^{-3/4}\, V_{\rm 4-cycle}\, \tau_{D7}^\pc\,,
\end{equation}
which may be compared with $\td\kappa=\sqrt{20R_5}\,c^{-3/4}$ coming from
(\ref{sigkap}).  Note in particular that the factor $c^{-3/4}$ is common
to both expressions.

    This in fact leads to the picture of a Randall-Sundrum
realization in the spirit of \cite{Chan:2000ms}, where the
Randall-Sundrum geometry arises from a warped F-theory construction.
In \cite{Chan:2000ms}, it was claimed that the six dimensions
transverse to the 3-brane may be viewed as being the base of an
elliptically fibered F-theory Calabi-Yau (complex) 4-fold.  Curvature
of the base may be related to the number of D3-branes and D7-branes present
as well as the topology of the 4-fold itself.  The kinked `thin-brane'
solution of (\ref{ads5met}) would then be an orbifold limit of this
F-theory construction.  Of course, the curvature sources arise not
from the orbifolding itself but rather from the D3-branes and D7-branes
pinned to the fixed orbifold planes.  There is nevertheless still a
need for an orbifold, since one needs to match not only the
delta-functions in the energy-momentum tensor but also in all the
form-field equations of motion.  In particular, a D7-brane would
require a charge for the 1-form $d\chi$, which however is absent in
the $Z_2$-identified flat space.  Thus this D7-brane charge would need
to be cancelled by something else, and an orbifold charge might be a
natural candidate for this.

   If this proposal were correct then our view of Kaluza-Klein
theories would have come full circle.  Initially it was thought that
extra dimensions had to be compact in order to ensure a proper
four-dimensional behavior.  Then Randall-Sundrum II proposed an
abandonment of this long-held view by demonstrating how an appropriate
warp factor in the metric is all that is needed for the binding of
modes to the brane.  But with the suggestion of \cite{Chan:2000ms}
that the space transverse to the D3-branes is simply the base of an
F-theory Calabi-Yau 4-fold (whether smooth or in the orbifold limit),
one once again returns to the original picture that a Kaluza-Klein
theory necessarily has a compact internal space.

\section*{Acknowledgements}

   M.C., C.N.P. and K.S.S. should like to thank SISSA for hospitality
during the early stages of this work.  M.C. should like to thank The
High Energy Theory Group at Rutgers University, for hospitality.  Work
supported in part by the European Commission RTN programme
HPRN-CT-2000-00131, and by the programme {\it Supergravity,
Superstrings and M-theory} of the Centre \'Emile Borel of the Institut
Henri Poincar\'e, Paris (UMS 839-CNRS/UPMC).

\end{document}